# Impact of Doping Spiro-OMeTAD with Li-TFSI, FK209, and tBP on the Performance of Perovskite Solar Cells

Mehran Hosseinzadeh Dizaj  *

*Department of Electrical engineering, Islamic Azad University, Central Tehran Branch, Tehran, Iran. E-mail: Mehran72002@gmail.com*

**Abstract:** This study delves into the effects of doping Spiro-OMeTAD, a widely employed hole transport material (HTM) in perovskite solar cells, with three distinct impurities: lithium bis(trifluoromethanesulfonyl)imide (Li-TFSI), FK209, and tert-butylpyridine (tBP). The focus lies on their influence on critical photovoltaic parameters, including Fill Factor ($FF$), power conversion efficiency ($\eta$), short-circuit current density ($Jsc$), and open-circuit voltage ($Voc$). Employing advanced simulation tools, we systematically compare the performance of undoped and doped Spiro-OMeTAD structures. Among the examined dopants, Li-TFSI exhibits the most pronounced enhancements, significantly improving both $FF$ and efficiency. In contrast, FK209 and tBP yield moderate gains in device performance. These results underscore the superior charge transport properties imparted by Li-TFSI doping and its capacity to optimize HTM functionality, thereby presenting a promising approach to advancing the efficiency and stability of perovskite solar cells.

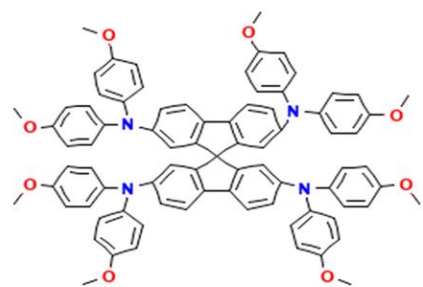

## 1. Introduction

Perovskite solar cells (PSCs) are rapidly gaining recognition as one of the most promising photovoltaic technologies due to their remarkable efficiency, cost-effective fabrication processes, and versatile device architectures.[1] A vital element in these devices is the hole transport material (HTM), which facilitates the extraction and transport of photogenerated holes to the electrode, ensuring efficient charge separation and collection.[2] Among HTMs, Spiro-OMeTAD has emerged as the most prevalent due to its optimal energy alignment with perovskite absorbers and its proven compatibility with high-efficiency device configurations. Nevertheless, the intrinsic conductivity and hole mobility of undoped Spiro-OMeTAD are relatively low, necessitating the addition of specific dopants to enhance its electrical properties. Key dopants such as lithium bis(trifluoromethanesulfonyl)imide (Li-TFSI), (FK209), and tert-butylpyridine (tBP) are frequently employed to improve the performance of Spiro-OMeTAD-based PSCs.[3] Doping plays a crucial role in optimizing the HTM layer by enhancing its charge carrier mobility and electrical conductivity, which directly affects key photovoltaic parameters such as open-circuit voltage ($Voc$), fill factor ($FF$), and power conversion efficiency (PCE). For example, Li-TFSI facilitates the chemical oxidation of Spiro-OMeTAD, significantly improving its conductivity and hole extraction efficiency. FK209, a cobalt-based complex, increases the density of charge carriers, thereby reducing charge recombination and enhancing charge transport.[4] Additionally, tBP, as a secondary additive, improves the morphological properties of the HTM layer, reducing interfacial resistance and stabilizing device performance. By incorporating these dopants, the photovoltaic performance of PSCs can be optimized, emphasizing the importance of tailoring HTM properties to achieve higher efficiencies and operational stability.[5]

In this work, we conduct a systematic investigation into the effects of doping Spiro-OMeTAD with Li-TFSI, FK209, and tBP, employing SCAPS (Solar Cell Capacitance Simulator) software to model and analyze their impact on charge carrier mobility and its correlation with critical photovoltaic parameters.[6] The study focuses on evaluating the influence of these dopants on fill factor ($FF$), power conversion efficiency (PCE), open-circuit voltage ($Voc$), and short-circuit current density ($Jsc$). By comparing the performance of pristine Spiro-OMeTAD with its doped counterparts, the results offer valuable insights into the role of dopants in improving HTM functionality and overall solar cell efficiency[7] This analysis aims to guide the design and development of high-performance PSCs, paving the way for their potential commercialization. The chemical structure of Spiro-OMeTAD is shown in Figure 1.[8]

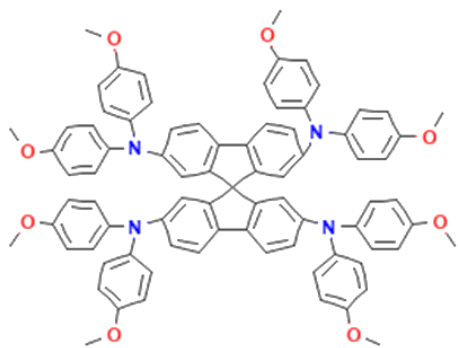

**Figure 1.** Spiro-OMeTAD Structure.

## 2. Methodology

In this research, we adopt the solar cell structure provided in Table 1 of a foundational study as the baseline for our simulations and parameter extraction. This structure, derived from the well-established architecture of thin-film inorganic semiconductor solar cells, provides a reliable framework for modeling perovskite solar cells.[9] Key parameters such as layer thickness, carrier mobilities, bandgaps, and energy band alignments are extracted from the reference structure and serve as the foundation for calibrating our device model.[10]

Using these initial parameters, we first simulate a perovskite solar cell employing a pristine Spiro-OMeTAD

HTM layer. This baseline model allows for accurate comparison with modified structures. Subsequently, the effects of doping Spiro-OMeTAD with Li-TFSI, FK209, and tBP are systematically introduced into the simulations. Each dopant's influence on critical parameters, such as electrical conductivity, mobility, recombination dynamics, and energy levels, is incorporated into the model.[11] By iteratively adjusting the device configuration, we assess the effects of each doping additive on fill factor (*FF*), open-circuit voltage (*Voc*), short-circuit current density (*Jsc*), and power conversion efficiency (PCE).[12]

This ste-by-step approach ensures consistency with prior validated models while providing robust insights into the role of HTM doping in optimizing solar cell performance. By leveraging the established structure as a reference, our simulations achieve improved accuracy and relevance, enabling a comprehensive analysis of the potential improvements brought about by these dopants.[13]

## 3. Laboratory Methods for Adding Impurities to Spiro-OMeTAD

To improve the performance of Spiro-OMeTAD as a hole transport material (HTM) in perovskite solar cells, specific dopants such as Li-TFSI, FK209, and tBP are introduced during the preparation of the HTM solution. Below is a detailed explanation of how these dopants are added in a laboratory setting:[14]

**Table 1.** Parameters used to simulate perovskite solar cells in SCAPS

| Definitions | Name | SnO₂( ETL) | Spiro-OMeTAD (HTL) | ILHTL & ILETL | PSC | TiO₂ |
|---|---|---|---|---|---|---|
| Thickness | t (nm) | 500 | 350 | 10 | 330 | 50 |
| Bandgap | Eg (eV) | 3.50 | 3 | 1.55 | 1.5 | 3.2 |
| Electron Affinity | $\mathcal{X}$ (eV) | 4 | 2.45 | 3.9 | 3.9 | 3.9 |
| Dielectric Permittivity | εr | 9 | 3 | 6.5 | 6.5 | 9 |
| CB effective DOS | NC (cm⁻³) | $2.2 \times 10^{18}$ | $2.2 \times 10^{18}$ | $2.2 \times 10^{18}$ | $2.2 \times 10^{18}$ | $2.2 \times 10^{18}$ |
| VB effective DOS | NV (cm⁻³) | $1.8 \times 10^{19}$ | $1.8 \times 10^{19}$ | $1.8 \times 10^{19}$ | $1.8 \times 10^{19}$ | $1.8 \times 10^{19}$ |
| Electron thermal velocity | Vth-e (cm/s) | $10^7$ | $10^7$ | $10^7$ | $10^7$ | $10^7$ |
| Hole thermal velocity | Vth-h (cm/s) | $10^7$ | $10^7$ | $10^7$ | $10^7$ | $10^7$ |
| Electron mobility | μe (cm²/Vs) | 20 | $2 \times 10^{-4}$ | 2 | 2 | 20 |
| Hole mobility | μh (cm²/Vs) | 10 | $2 \times 10^{-4}$ | 2 | 2 | 10 |
| Acceptor Concentration | NA (cm⁻³) | - | $10^{19}$ | $10^{13}$ | $10^{13}$ | - |
| Donor Concentration | ND (cm⁻³) | $2 \times 10^{19}$ | - | $10^{13}$ | $10^{13}$ | $2 \times 10^{19}$ |
| Defect Density | Nt (cm⁻³) | $10^{15}$ | $10^{15}$ | $10^{17}$ | $2.5 \times 10^{13}$ | $10^{15}$ |
| Electron cross capture | σe (cm²) | $2 \times 10^{-14}$ | $2 \times 10^{-14}$ | $2 \times 10^{-14}$ | $2 \times 10^{-14}$ | $2 \times 10^{-14}$ |
| Hole cross capture | σh (cm²) | $2 \times 10^{-14}$ | $2 \times 10^{-14}$ | $2 \times 10^{-14}$ | $2 \times 10^{-14}$ | $2 \times 10^{-14}$ |

**Table 2.** Parameters changed in the HTM layer for each of the elements used as doping for the perovskite solar cell simulation in SCAPS

| Parameters | Spiro-OMeTAD | Li-TFSI | FK209 | tBP |
|---|---|---|---|---|
| Electrical Conductivity ($S/cm$) | 0.0002 | 1e-06 | 1e-05 | 1e-05 |
| Hole Mobility ($cm^2$) | 0.0002 | 0.0002 | 0.0005 | 0.0001 |
| Effective Density of States in the Valence Band, Nv ($cm^{-3}$) | 1e+18 | 1e+18 | 1e+18 | 1e+18 |
| Recombination Coefficient ($cm^3$) | 1e-14 | 1e-11 | 1e-10 | 1e-10 |
| Fermi Level ($eV$) | 0.5 | 0.4 | 0.3 | 0.3 |
| Layer Thickness (nm) | 300.0 | 300.0 | 300.0 | 300.0 |
| Relative Permittivity | 3.0 | 4.0 | 3.0 | 3.0 |

### A. Adding Li-TFSI (Lithium Bis(trifluoromethanesulfonyl)imide):

1. **Preparation:** A stock solution of Li-TFSI is prepared by dissolving a precise amount (e.g., 520 mg/mL) in acetonitrile, a polar solvent chosen to ensure full dissolution.
2. **Integration**: After preparing the Spiro-OMeTAD solution in chlorobenzene at a typical concentration of 72.3 mg/mL, a calculated volume of the Li-TFSI solution is added.[15]
3. **Mixing:** The resulting mixture is stirred thoroughly at room temperature to achieve uniform distribution of Li-TFSI in the solution.
4. **Effect:** Li-TFSI enhances the oxidation state of Spiro-OMeTAD, significantly improving its electrical conductivity and hole mobility. The chemical structure of Li-TFSI is shown in Figure 2.

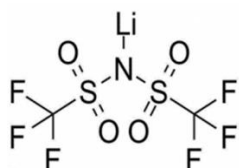

**Figure 2.** Li-TFSI Structure.

### B. Adding FK209 (Cobalt Complex):

1. **Preparation:** FK209 is dissolved in acetonitrile to achieve a concentration of approximately 300 mg/mL, though this may vary depending on the desired doping level.
2. **Integration:** The FK209 solution is precisely measured and added to the Spiro-OMeTAD solution alongside Li-TFSI.[16]
3. **Mixing:** The solution is gently stirred to ensure complete dissolution of FK209 and its homogeneous incorporation into the Spiro-OMeTAD matrix.
4. **Effect:** FK209 enhances charge carrier density, reduces recombination losses at the HTM/perovskite interface, and improves the device's overall stability. The chemical structure of FK209 is shown in Figure 3.[17]

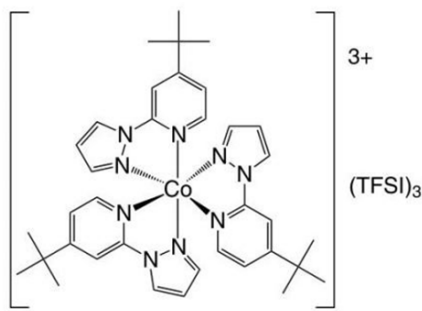

**Figure 3.** FK209 Structure.

### C. Adding tBP (Tertiary Butylpyridine):

1. **Preparation:** Unlike Li-TFSI and FK209, tBP is a liquid additive that does not require prior dissolution.
2. **Integration:** A small volume of tBP (e.g., 30 μL per mL of Spiro-OMeTAD solution) is added directly after the inclusion of Li-TFSI and FK209.[18]
3. **Mixing:** The solution is stirred gently to ensure homogeneity.
4. **Effect:** tBP improves the morphology of the Spiro-OMeTAD layer by reducing pinholes and optimizing the energy alignment between the HTM and the perovskite absorber. The chemical structure of tBP is shown in Figure 4.

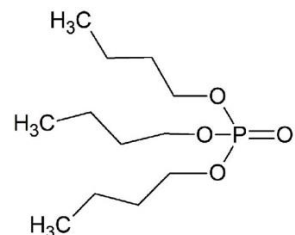

**Figure 4.** tBP Structure.

### Deposition of the HTM Layer:

Once the doped Spiro-OMeTAD solution is prepared, it is deposited onto the perovskite layer via spin-coating. The coated film is subsequently oxidized under ambient atmospheric conditions, which further enhances its conductivity and prepares it for the deposition of the top electrode, such as gold.[19]

### Simulation-Based Approach:

In this study, we do not carry out the physical doping or deposition processes. Instead, we simulate the effects of these dopants computationally using SCAPS (Solar Cell Capacitance Simulator) software.[20] The changes induced by Li-TFSI, FK209, and tBP are incorporated into the model by modifying key HTM parameters, including:

- Electrical Conductivity,
- Hole Mobility,
- Recombination Coefficient, and
- Effective Density of States in the Valence Band ($N_v$).

This simulation allows us to evaluate the influence of these dopants on critical photovoltaic metrics such as fill factor ($FF$), power conversion efficiency ($PCE$ or $\eta$), open-circuit voltage ($V_{oc}$), and short-circuit current density ($J_{sc}$) without requiring physical synthesis or experimentation.[21] The computational approach provides a cost-effective and efficient way to assess the potential improvements brought by each dopant, guiding future experimental efforts. The current-voltage and quantum efficiency diagrams are shown in Figures 5 and 6.[22]

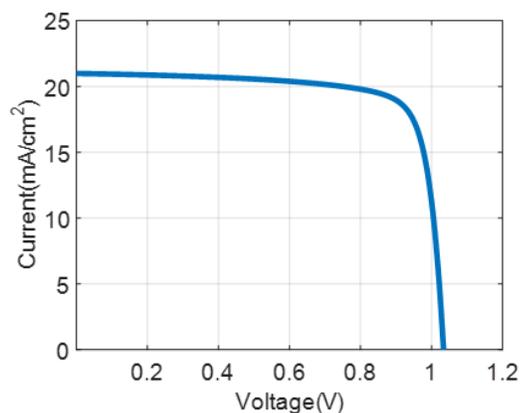

**Figure 5.** Voltage-current diagram of a basic solar cell.

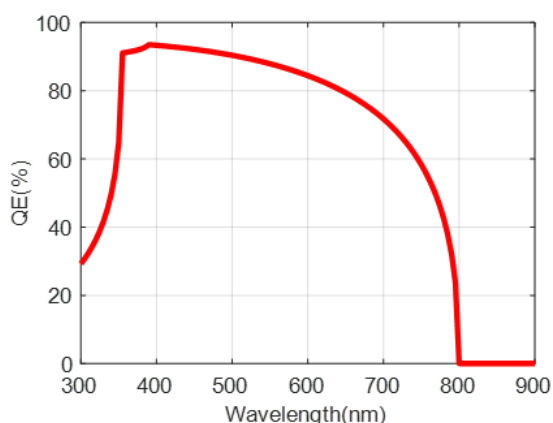

**Figure 6.** Basic solar cell quantum efficiency chart.

**Simulation of Doping Effects in the HTM Layer:**
In the initial phase of this study, the effects of dopants, including Li-TFSI, FK209, and tBP, are incorporated into the Spiro-OMeTAD hole transport material (HTM) layer via parameter modification in simulations, eliminating the need to introduce new physical layers. Key intrinsic properties of the Spiro-OMeTAD layer are adjusted to model the chemical and electrical influences of the dopants.[23]

**Key Adjustments:**
1. **Electrical Conductivity:**
The electrical conductivity is enhanced to reflect the improved charge transport properties enabled by the dopants. This parameter accounts for the dopant-induced oxidation of Spiro-OMeTAD, which significantly increases the mobility and density of charge carriers.

2. **Hole Mobility:**
Adjustments to hole mobility simulate the enhanced efficiency of hole extraction and transport in the doped HTM, a critical factor for improved photovoltaic performance.

3. **Effective Density of States in the Valence Band (Nv):**
The Nv parameter is modified to reflect the increased availability of charge carriers due to the chemical effects of the dopants, which improve charge injection and carrier population.

4. **Recombination Coefficient:**
A lower recombination coefficient is introduced to represent the suppression of carrier recombination. This adjustment is indicative of better interface properties, reduced defect densities, and the dopants' positive impact on the quality of the HTM layer.

5. **Fermi Level Position:**
Shifts in the Fermi level simulate changes in charge carrier concentration and energy level alignment within the device, ensuring an optimized interface with adjacent layers.

6. **Relative Permittivity:**
Adjustments to the relative permittivity represent changes in the dielectric constant caused by the dopants, simulating their influence on the layer's polarization properties.
Invariance of Structural Dimensions:
The layer thickness of the Spiro-OMeTAD remains unchanged, as doping does not directly alter the physical dimensions of the HTM layer. Instead, the modifications are confined to the material's intrinsic properties.[24]

**Simulation Implementation:**
The above parameter modifications are integrated into the SCAPS (Solar Cell Capacitance Simulator) software to comprehensively assess the dopants' impact on photovoltaic performance metrics, including:
- Fill Factor ($FF$),
- Power Conversion Efficiency ($PCE$ or $\eta$),
- Open-Circuit Voltage ($Voc$), and
- Short-Circuit Current Density ($Jsc$).

By focusing on intrinsic parameter adjustments, this computational approach offers an efficient and systematic method to explore the doping effects on the HTM layer. It enables accurate modeling of the dopants' influence on Spiro-OMeTAD, streamlining the evaluation process while minimizing computational complexity. This methodology provides valuable insights into the mechanisms underlying performance enhancement in perovskite solar cells.[25]

**Simulation Results for Case 1: Doping Spiro-OMeTAD with Li-TFSI**
Doping Spiro-OMeTAD with Li-TFSI demonstrates measurable improvements in the performance of perovskite solar cells (PSCs). Below is a detailed comparison of the key photovoltaic parameters before and after doping, highlighting the advantages and mechanisms of Li-TFSI's influence.

**Key Simulation Results:**

**1. Fill Factor (*FF*):**
- Before Doping: 78.77%
- After Doping: 79.68%
- Improvement: 1.15%

Li-TFSI doping enhances the charge transport and reduces resistive losses in the HTM, leading to more efficient power transfer to the external circuit.

**2. Efficiency (*η*):**
- Before Doping: 17.09%
- After Doping: 17.32%
- Improvement: 1.35%

The improvement in efficiency is primarily due to enhanced charge carrier mobility and conductivity, which reduce recombination losses and enable better charge extraction.

**3. Short-Circuit Current Density (*Jsc*):**
- Before Doping: 20.988 mA/cm²
- After Doping: 21.004 mA/cm²
- Improvement: 0.08%

The slight increase in *Jsc* suggests that Li-TFSI doping minimally affects photogeneration but improves the efficiency of charge collection, contributing to the enhanced overall performance.

**4. Open-Circuit Voltage (*Voc*):**
- Before Doping: 1.0338 V
- After Doping: 1.0347 V
- Improvement: 0.087%

Li-TFSI doping results in a marginal increase in *Voc* by improving energy alignment between the HTM and perovskite layers and reducing recombination losses.

**Advantages of Li-TFSI Doping**

1. **Enhanced Conductivity:**
Li-TFSI promotes the oxidation of Spiro-OMeTAD, leading to higher electrical conductivity. This enhancement reduces series resistance within the HTM, enabling more efficient charge transport.

2. **Reduced Recombination Losses:**
The doping reduces recombination at the HTM/perovskite interface by stabilizing the HTM layer and improving charge carrier mobility. This contributes significantly to the observed improvements in FF and efficiency.

3. **Marginal Gains in *Voc* and *Jsc*:**
While the increases in *Voc* and Jsc are modest, they collectively improve the overall device performance. These gains reflect the optimized energy alignment and improved charge extraction facilitated by the doping.

**4. Simplicity of Process:**
The addition of Li-TFSI involves straightforward solution processing and does not require complex changes in fabrication methods, making it an accessible and scalable strategy for enhancing PSC performance. The current-voltage and quantum efficiency diagrams of Li-TFSI doping are shown in Figures 7 and 8.[22]

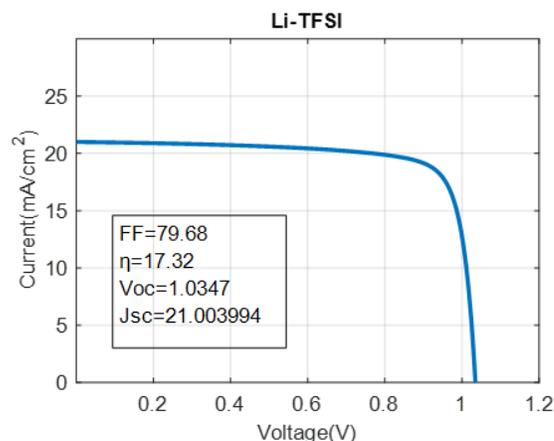

**Figure 7.** Current-voltage diagram of Li-TFSI-doped solar cell in HTM.

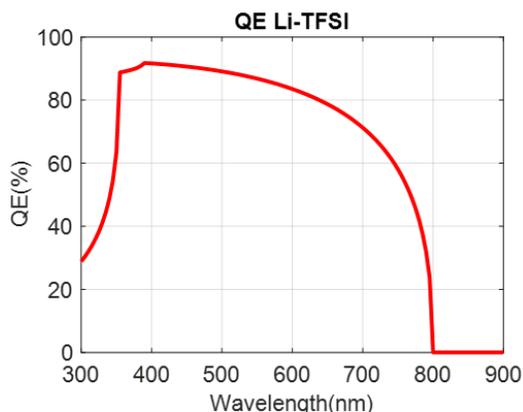

**Figure 8.** Quantum efficiency diagram of Li-TFSI-doped solar cell in HTM.

**Simulation Results for Case 2: Doping Spiro-OMeTAD with FK209**

Doping Spiro-OMeTAD with FK209, a cobalt-based complex, shows improvements in the performance of perovskite solar cells (PSCs), albeit with slightly less pronounced gains compared to Li-TFSI doping. Below is a detailed breakdown of the key photovoltaic parameter changes and the implications of FK209 doping.

**Key Simulation Results**

**1. Fill Factor (*FF*):**
- Undoped Spiro-OMeTAD: 78.77%
- Doped with FK209: 79.41%
- Improvement: 0.81%

FK209 doping enhances charge transport by increasing the conductivity and mobility of the HTM, reducing resistive losses and improving the $FF$.

**2. Efficiency ($\eta$):**
- Undoped Spiro-OMeTAD: 17.09%
- Doped with FK209: 17.26%
- Improvement: 1.00%

The improved efficiency reflects FK209's ability to facilitate better charge carrier extraction and reduce energy losses, resulting in more effective power conversion.

**3. Short-Circuit Current Density ($Jsc$):**
- Undoped Spiro-OMeTAD: 20.988 mA/cm²
- Doped with FK209: 21.002 mA/cm²
- Improvement: 0.07%

The minor increase in $Jsc$ suggests that FK209 doping does not significantly affect photogeneration but slightly enhances charge collection efficiency.

**4. Open-Circuit Voltage ($Voc$):**
- Undoped Spiro-OMeTAD: 1.0338 V
- Doped with FK209: 1.0347 V
- Improvement: 0.087%

The marginal increase in $Voc$ indicates reduced recombination losses and improved energy alignment between the HTM and adjacent layers due to FK209 incorporation.

**Advantages of FK209 Doping**

1. **Enhanced $FF$ and Efficiency:**
FK209 doping improves $FF$ and efficiency, demonstrating its effectiveness in enhancing charge transport and minimizing energy losses, though the gains are slightly smaller than those observed with Li-TFSI doping.

**2. Improved Charge Transport and Stability:**
- FK209 increases the density of charge carriers in Spiro-OMeTAD, boosting hole extraction and transport efficiency.
- It also stabilizes the HTM layer, potentially enhancing the device's operational stability over time, a critical consideration for commercial applications.

3. **Reduced Recombination:**
The doping reduces recombination at the HTM/perovskite interface, leading to improved device performance.

4. **Marginal Changes in $Voc$ and $Jsc$:**
As with Li-TFSI doping, the effects of FK209 on $Voc$ and $Jsc$ are minimal, reinforcing the understanding that its primary role is in optimizing charge transport rather than affecting photogeneration or photovoltage. The current-

voltage diagram and quantum efficiency of FK209 doping are seen in Figures 9 and 10.[1]

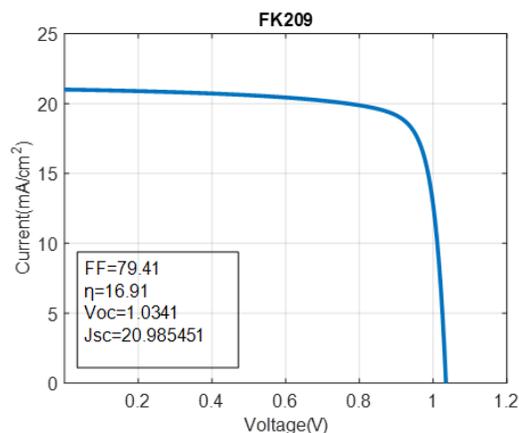

**Figure 9.** Current-voltage diagram of FK209-doped solar cell in HTM.

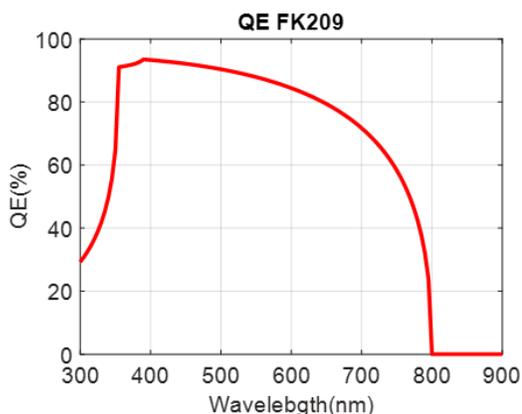

**Figure 10.** Quantum efficiency diagram of FK209-doped solar cell in HTM.

**Simulation Results for Case 3: Doping Spiro-OMeTAD with tBP**
Doping Spiro-OMeTAD with tBP (tertiary butylpyridine) demonstrates mixed effects on the performance of perovskite solar cells (PSCs). While tBP is known for its ability to improve film morphology and reduce defects, its impact on key photovoltaic parameters in this simulation is limited and, in some cases, detrimental. Below is a detailed analysis of the results and their implications.[26]

**Key Simulation Results**

1. **Fill Factor ($FF$):**
- Undoped Spiro-OMeTAD: 78.77%
- Doped with tBP: 77.96%
- Change: Decreased by 1.03%

The decrease in $FF$ suggests that tBP slightly increases resistive losses or introduces disruptions in charge transport, possibly due to changes in film conductivity despite improvements in morphology.

**2. Efficiency ($\eta$):**
- Undoped Spiro-OMeTAD: 17.09%
- Doped with tBP: 16.93%
- Change: Decreased by 0.94%

The reduction in efficiency indicates that the improvements in interface quality or morphology brought by tBP are not sufficient to counteract the adverse effects on charge transport and recombination dynamics.

**3. Short-Circuit Current Density ($Jsc$):**
- Undoped Spiro-OMeTAD: 20.988 mA/cm²
- Doped with tBP: 20.993 mA/cm²
- Change: Increased by 0.02%

The negligible increase in $Jsc$ suggests that tBP has minimal influence on photogenerated current or charge collection efficiency.

**4. Open-Circuit Voltage ($Voc$):**
- Undoped Spiro-OMeTAD: 1.0338 V
- Doped with tBP: 1.0347 V
- Change: Increased by 0.087%

The slight improvement in $Voc$ reflects better energy alignment at the HTM/perovskite interface and a possible reduction in recombination losses, though the impact is modest.

**Overall Effects of tBP Doping**

1. **Film Morphology Improvements:**
tBP is well-documented to improve the uniformity and quality of the HTM layer by reducing pinholes and defects. These morphological benefits can enhance interface stability and reduce localized recombination. However, in this case, these advantages do not translate into significant gains in electrical performance.

2. **Minimal Impact on $Jsc$:**
The negligible change in $Jsc$ implies that tBP does not significantly influence light absorption or photogenerated charge carrier dynamics. Its role appears to be limited to morphology and interface adjustments rather than bulk transport properties.

3. **Reduction in $FF$ and Efficiency:**
The decrease in FF and efficiency highlights a trade-off between morphological improvements and adverse effects on charge transport properties. This could stem from an increase in resistive losses or suboptimal conductivity changes in the HTM layer induced by tBP.

4. **Slight $Voc$ Enhancement:**
The minor increase in $Voc$ suggests that tBP has a stabilizing effect on the HTM layer and improves energy alignment, leading to reduced recombination at the interface. However, the impact is relatively small compared to dopants like Li-TFSI or FK209. The current-voltage diagram and quantum efficiency of tBP doping are seen in Figures 11 and 12.[27]

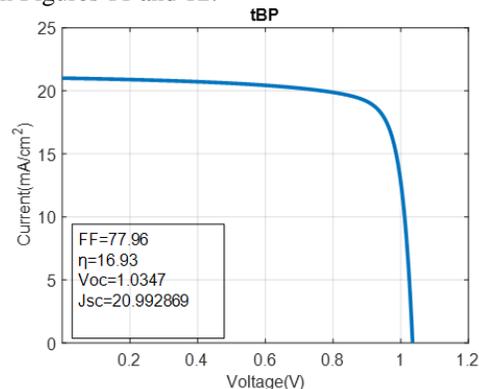

**Figure 11.** Current-voltage diagram of tBP-doped solar cell in HTM.

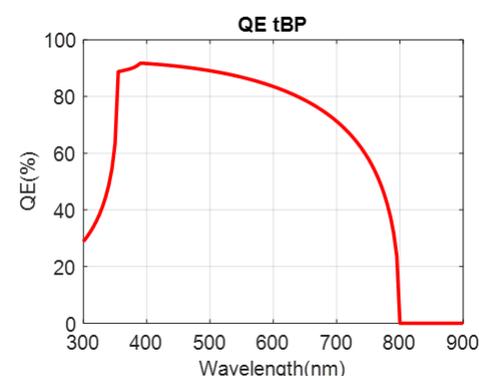

**Figure 12.** Quantum efficiency diagram of tBP-doped solar cell in HTM.

## 5. Result

**Table 3.** Comparison table of obtained parameters

| Doping Type | Fill Factor ($FF$) [%] | Efficiency ($\eta$) [%] | Short-Circuit Current Density ($Jsc$) [mA/cm²] | Open-Circuit Voltage ($Voc$) [V] |
|---|---|---|---|---|
| Spiro-OMeTAD | 78.77 | 17.09 | 20.988 | 1.0338 |
| Li-TFSI | 79.68 | 17.32 | 21.004 | 1.0347 |
| FK209 | 79.41 | 17.26 | 21.002 | 1.0347 |
| tBP | 77.96 | 16.93 | 20.993 | 1.0347 |

**Comparative Analysis of Spiro-OMeTAD Doping with Li-TFSI, FK209, and tBP**

The simulation-based evaluation of doping Spiro-OMeTAD with Li-TFSI, FK209, and tBP reveals significant differences in their effects on the key photovoltaic parameters of perovskite solar cells (PSCs). The undoped Spiro-OMeTAD serves as the baseline, and the relative impacts of these dopants on fill factor ($FF$), efficiency ($\eta$), short-circuit current density ($Jsc$), and open-circuit voltage ($Voc$) are outlined below.

**1. Li-TFSI: Most Effective Performance Enhancer**
- **Fill Factor ($FF$):** Increased from 78.77% to 79.68% (+1.15%).

- **Efficiency ($\eta$):** Improved from 17.09% to 17.32% (+1.35%).
- **Short-Circuit Current Density ($Jsc$):** Slight rise from 20.988 mA/cm² to 21.004 mA/cm² (+0.08%).
- **Open-Circuit Voltage ($Voc$):** Marginal increase from 1.0338 V to 1.0347 V (+0.087%).

**Interpretation:**

Li-TFSI demonstrates the most pronounced improvement in $FF$ and efficiency, reflecting its ability to significantly enhance the charge transport properties of Spiro-OMeTAD. Its effects on $Jsc$ and $Voc$ are smaller but still positive, indicating reduced recombination losses and better energy alignment. These results confirm Li-TFSI's role as a highly effective dopant for optimizing overall solar cell performance.

**2. FK209: Noticeable but Lesser Gains**

- **Fill Factor ($FF$):** Increased from 78.77% to 79.41% (+0.81%).
- **Efficiency ($\eta$):** Improved from 17.09% to 17.26% (+1.00%).
- **Short-Circuit Current Density ($Jsc$):** Slight rise from 20.988 mA/cm² to 21.002 mA/cm² (+0.07%).
- **Open-Circuit Voltage ($Voc$):** Marginal increase from 1.0338 V to 1.0347 V (+0.087%).

**Interpretation:**

Doping with FK209 produces improvements in FF and efficiency comparable to those seen with Li-TFSI, though the gains are slightly smaller. FK209 primarily enhances charge transport and reduces recombination losses in the HTM layer, without significantly influencing carrier generation or photovoltage. While its effects are less impactful than Li-TFSI, FK209 still represents a valuable option for performance enhancement.

**3. tBP: Limited Performance Gains but Auxiliary Benefits**

- **Fill Factor ($FF$):** Decreased from 78.77% to 77.96% (-1.03%).
- **Efficiency ($\eta$):** Dropped from 17.09% to 16.93% (-0.94%).
- **Short-Circuit Current Density ($Jsc$):** Minimal rise from 20.988 mA/cm² to 20.993 mA/cm² (+0.02%).
- **Open-Circuit Voltage ($Voc$):** Marginal increase from 1.0338 V to 1.0347 V (+0.087%).

**Interpretation:**

Unlike Li-TFSI and FK209, tBP does not improve $FF$ or efficiency. The slight reduction in $FF$ suggests increased resistive losses or suboptimal conductivity, while the negligible changes in $Jsc$ and $Voc$ confirm tBP's limited direct impact on electrical performance. However, tBP is known to improve HTM layer morphology and reduce interfacial defects, contributing to long-term device stability and complementary benefits when used alongside other dopants.

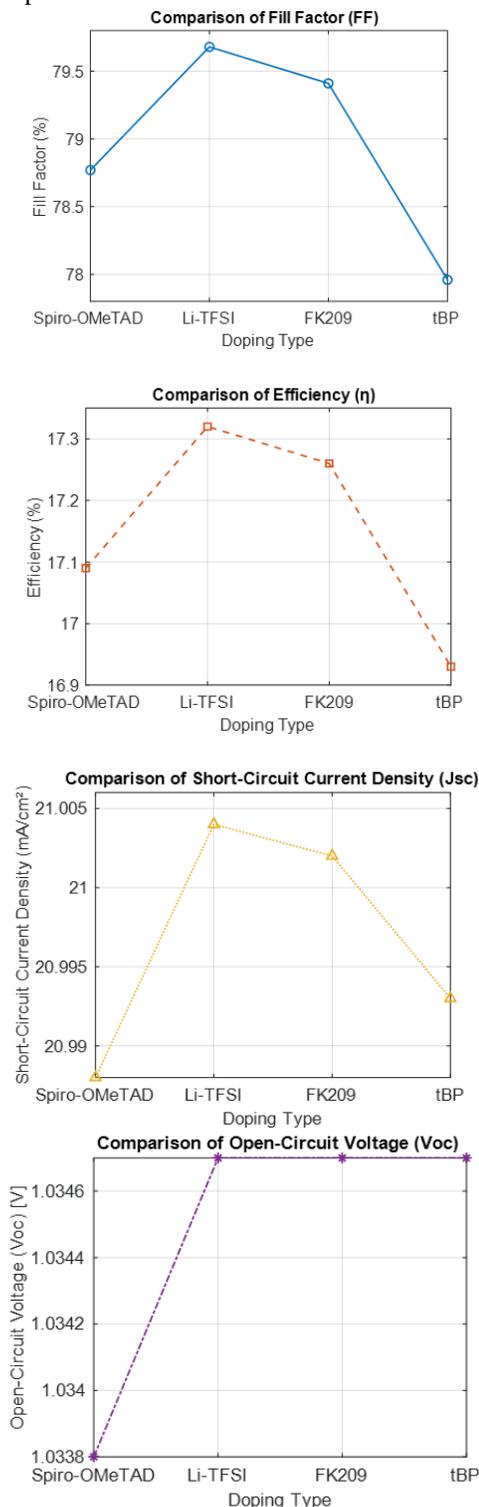

**Figure 13.** Comparative diagrams of solar cell parameters in the ground state and all 3 doped states.

## 6. Conclusion

The results of doping Spiro-OMeTAD with Li-TFSI, FK209, and tBP show notable improvements in certain photovoltaic parameters, particularly in Fill Factor (*FF*) and Efficiency (*η*). The doping with Li-TFSI yielded the highest improvements, enhancing *FF* to 79.68% and efficiency to 17.32%, compared to the undoped Spiro-OMeTAD (*FF* = 78.77%, *η* = 17.09%). Similarly, doping with FK209 and tBP resulted in minor improvements in *FF* and efficiency, though the changes were less pronounced. *Jsc* values remained relatively constant across all doping types, while *Voc* showed negligible variation, indicating that the primary impact of doping was on charge transport and recombination dynamics rather than the open-circuit voltage. Overall, the use of Li-TFSI as a dopant offers the most significant enhancement in the device performance, suggesting its potential as an effective doping agent for improving the efficiency of Spiro-OMeTAD-based perovskite solar cells.

## 7. References

1. T. Minemoto, M. Murata, *J. Appl. Phys.*, **2014**, 116, 054505.

2. M. Elawad, A. A. Elbashir, M. Sajid, K. I. John, H. Nimir, L. Yang, A. K. Ziyada, A. Osman, F. Rajab, *J. Chem. Phys.*, **2024**, 160, 444707.

3. X. Geng, M. Abdellah, R. Bericat Vadell, M. Folkenant, T. Edvinsson, J. Sá, *Nanomaterials*, **2021**, *11*, 3329.

4. X. Zhang, Y. Wang, G. Li, L. Huang, J. Yang, X. Qiu, W. Sun, *J. Phys. Chem. C*, **2022**, *126*, 22, 9528-9540.

5. G. Li, Y. Wang, L. Huang, W. Sun, *J. Alloys Compd.*, **2022**, *907*, 164432.

6. A. Pellaroque, N. K. Noel, S. N. Habisreutinger, Y. Zhang, S. Barlow, S. R. Marder, H. J. Snaith, *ACS Energy Lett.*, **2017**, *2*, 2044-2050.

7. Y. Shen, K. Deng, L. Li, *Small Methods*, **2022**, *6*, 11, 2200757.

8. A. Mutlu, D. Çırak, T. Yeşil, C. Zafer, B. Gultekin, *Org. Electron.*, **2023**, *113*, 106674.

9. J. Wang, J. Zhang, Y. Yang, Y. Dong, W. Wang, B. Hu, J. L, W. Cao, K. Lin, D. Xia, R. Fan, *Chem. Eng. J.*, **2022**, *429*, 132481.

10. J. Xu, P. Shi, K. Zhao, L. Yao, C. Deger, S. Wang, X. Zhang, S. Zhang, Y. Tian, X. Wang, J. Shen, C. Zhang, I. Yavuz, J. Xue, R. Wang, *ACS Energy Lett.*, **2024**, *9*, 1073-1081.

11. Y. Han, G. Zhang, H. Xie, T. Kong, Y. Li, Y. Zhang, J. Song, D. Bi, *Nano Energy*, **2022**, *96*, 107072.

12. Z. Wan, S. Jiang, H. Lu, J. Zhu, Y. Wang, H. Zeng, H. Yin, R. Wei, J. Luo, C. Jia, *J. Mater. Chem. C*, **2024**, *12*, 22, 8078-8086.

13. Y. Liu, Y. Hu, X. Zhang, P. Zeng, F. Li, B. Wang, Q. Yang, M. Liu, *Nano Energy*, **2020**, *70*, 104483.

14. X. Guo, J. Li, B. Wang, P. Zeng, F. Li, Q. Yang, Y. Chen, M. Liu, *ACS Appl. Energy Mater.*, **2019**, *3*, 970-976.

15. M. Namatame, M. Yabusaki, T. Watanabe, Y. Ogomi, S. Hayase, K. Marumoto, *Appl. Phys. Lett.*, **2017**, *110*, 12.

16. F. Lin, J. Luo, Y. Zhang, J. Zhu, H. Ashraf, Z. Wan, C. Jia, *J. Mater. Chem. A*, **2023**, *11*, 6, 2544-2567.

17. S. Wang, Z. Huang, X. Wang, Y. Li, M. Günther, S. Valenzuela, P. Parikh, A. Cabreros, W. Xiong, Y. S. Meng, *J. Am. Chem. Soc.*, **2018**, *140*, 48, 16720-16730.

18. E. J. Juarez-Perez, M. R. Leyden, S. Wang, L. K. Ono, Z. Hawash, Y. Qi, *Chem. Mater.*, **2016**, *28*, 16, 5702-5709.

19. J. Zhang, Z. Li, C. Li, L. Mu, S. Tang, K. Cai, Z. Cheng, C. Liu, S. Xiang, Z. Zhang, *Chinese Chem. Lett.*, **2024**, 110046.

20. F. Lamberti, T. Gatti, E. Cescon, R. Sorrentino, A. Rizzo, E. Menna, G. Meneghesso, M. Meneghetti, A. Petrozza, L. Franco, *Chem.*, **2019**, *5*, 1806-1817.

21. G. Ren, W. Han, Y. Deng, W. Wu, Z. Li, J. Guo, H. Bao, C. Liu, W. Guo, *J. Mater. Chem. A*, **2021**, *9*, 4589-4625.

22. D. Stanić, V. Kojić, T. Čižmar, K. Juraić, L. Bagladi, J. Mangalam, T. Rath, A. Gajović, *Materials* **2021**, 14, 6341.

23. L. Nakka, Y. Cheng, A. G. Aberle, F. Lin, *Adv. Energy Sustain. Res.*, **2022**, *3*, 2200045.

24. T. M. Vo, T. M. H. Nguyen, R. He, C. W. Bark, *ACS Omega*, **2024**, *9*, 46030-46040

25. L. Nakka, Y. Cheng, A. G. Aberle, F. Lin, **2022**, *3*, 2200045.

26. M. H. Dizaj, M. J. Torkamani, *Clin. Cancer Investig. J.*, **2023**, *11*, 1s.

27. V. Nazerian, M. Hosseinzadeh Dizaj, A. Assari, S. C. Shishvan, F. Shahnavaz, T. Sutikno, *TELKOMNIKA Telecommunication Computing Electronics and Control*, **2024**, *22*, 1293-1301.